# Digital Holographic Interferometry for Micro-Deformation Analysis of Morpho Butterfly Wing

Ali Mardan Dezfouli,[1] Nazif Demoli[1], Denis Abramović,[1,2] Mario Rakić[1], Hrvoje Skenderović[1,2]

[1]*Institute of Physics, Bijenička cesta 46, Zagreb 10000, Croatia*

[2]*Centre of Excellence for Advanced Materials and Sensing Devices, Photonics and Quantum Optics Unit, Institute Ruđer Bošković, Bijenička cesta 54, Zagreb 10000, Croatia*

*\*Corresponding author:* amdezfouli@ifs.hr

**Abstract:**

In this study, we present a detailed analysis of deflections in a butterfly wing utilizing digital holographic interferometry. Our methodology revolves around an off-axis lensless Fourier holographic setup, and we employ laser excitation to induce deflections in the object. The implementation of a digital holographic interferometry setup, tailored for rapid monitoring of micro-deformation, is a central aspect of the research. We offer an overview of the theoretical foundations of this technique, complemented by both experimental and simulated tests aimed at validating our findings. Significantly, our investigation focuses on the detailed analysis of the micro structures found on the wing of the Morpho butterfly. The insights garnered from our study not only affirm the precision and potential of this methodology but also shed light on promising avenues for further exploration, especially in the domain of high-precision deflection sensing and its diverse applications.

______________________________________________________________________

## Introduction

Engineered nanostructures often draw inspiration from the intricate nanoscale designs found in nature, which have evolved over billions of years to serve specific functions. In the realm of photonic applications, there is considerable fascination with the structural coloration exhibited by Morpho butterfly wings. Unlike traditional pigmentation, structural coloration relies on constructive interference of visible light. The small scales on Morpho butterfly wings possess ribbed lamellae layers with remarkable periodicity. This nanostructure selectively interacts with specific wavelengths of light, giving rise to the iconic blue iridescence synonymous with the Morpho butterfly. Beyond their aesthetic appeal, the hierarchical nanostructures of butterfly wings offer compelling potential for functional applications. These wings have found utility in a wide array of fields, including optical gas sensors [1], solar cells [2], infrared detectors [3], photocatalysts [4], surface-enhanced Raman spectroscopy (SERS) substrates [5], and more. Radiation detection attracts special attention as the need for low-cost imaging detectors with wide spectral ranges grows ever stronger [6].

In a previous study [7,8], digital holography was applied to explore the potential of utilizing Morpho menelaus butterfly wings as infrared (IR) detectors. The study delved into the biological structures of butterfly wings and how they exhibit potential for IR detection by undergoing spectral changes induced by thermal effects. The irradiation process and determination of the detection threshold for IR were elaborated.

We present here a more thorough examination of the observed thermal dilatation and bending of the wing scales in the context of radiation detection. One critical aspect of these nanostructures is the phase information embedded within them. The phase information relates to the arrangement and orientation of the microstructures, determining the structural color observed. In the context of radiation detection, understanding how phase information influences the mechanical deformation of wing scales is significant.

Our current research specifically focuses on conducting an in-depth analysis of the phase map to facilitate a thorough deformation analysis. Digital holographic interferometry (DHI) is a good choice for studying changes as it provides the advantages of high resolution, non-contact operation, and full-field imaging. Because of the availability of high-resolution CCD cameras and large-storage-capacity computers, DHI has replaced the lengthy procedure of wet processing of photographic storage materials and optical reconstruction. Using DHI, not only one can reconstruct the amplitude, but also the important phase can be reconstructed simultaneously. The DHI process involves recording two digital holograms: one captures the reference state of the object, while the other records the altered state of the object. These holograms are captured electronically using imaging sensors like CCD or CMOS, serving as reference and object holograms, respectively.

To extract information about the phase of the object's wavefront in different states, numerical reconstruction is performed on these digital holograms. The interference phase is obtained by subtracting the phase of the reference hologram (prior deformation state) from the phase of the object hologram (post-deformation state), all accomplished without the need for additional phase-shifting interferometry. In essence, DHI eliminates the necessity for extra steps in calculating the interference phase while encoding displacement information in the phase of an interference field signal. Due to its capacity to measure displacements at the micrometer or even submicrometer scale, Digital Holographic Interferometry (DHI) has found widespread application across various domains of research and industry[9]. Its versatility has been harnessed in diverse fields including vibroacoustics [10], optical metrology [11], flow analysis [12], and temperature measurements [13], among others.

In this research, we present the utilization of lensless Fourier transform digital holographic interferometry to evidence small displacements or deformations between an initial and a final (or deformed) state on Morpho butterfly wings. We have opted for this approach due to its simplicity in reconstruction techniques, ease of implementation, and reduced reliance on optical components. This technique enables the simultaneous observation of both real and virtual images. Its primary advantage lies in its capacity to be implemented through a single Fourier transformation of the hologram's transmittance, whereas alternative geometries necessitate multiple Fourier transforms or multiplication with an exponential factor.

In general, hologram reconstruction methods involve several fast Fourier transforms (FFT) and complex-matrix multiplications. Therefore, even in the case of high-speed digital processing, algorithms should be as simple and fast to compute as possible. The most used hologram reconstruction methods are the Fresnel and convolution approaches. Both of these methods allow extracting phase information and generating an interference phase map (or interferogram) to evidence small displacements or deformations between an initial and a final (or deformed) state [14]. In this work, we have analyzed small deflections of a morpho butterfly wing while it was being excited by a laser beam. The observed deflections were in the order of a few micrometers. Holograms were recorded in an standard off-axis digital lensless Fourier holographic setup. Subsequently, we employed the Fresnel reconstruction to derive the reconstruction algorithm to obtain the interference phase maps of the deformation created by laser excitation. First, we briefly discuss the theoretical aspects of the technique and present the experimental details. Finally, we show the experimental and simulated results which allowed us to perform a quantitative analysis of the induced deflection. These results set the basis for further research concerning the development of artificial photonics structures which could be used for radiation detection.

**Hologram Reconstruction**

The method involves capturing two digital holograms, representing the object in its original and deformed state. Employing the Fresnel method and a standard off-axis lensless digital Fourier holographic setup, each hologram is individually reconstructed using a single Fast Fourier Transform. In the Fresnel approximation (for both x- and y-values as well as for ξ- and η-values, which are small compared to the distance d between the reconstruction plane and the CCD), the complex amplitude is derived from the recorded hologram through a numerical reconstruction process, achieved by [15]:

$$\Gamma(\xi,\eta) = \frac{i}{\lambda d} \exp\left(\frac{-i2\pi d}{\lambda}\right) \exp\left[-i\frac{\pi}{\lambda d}\left(\xi^2 + \eta^2\right)\right]$$
$$\times \int_{-\infty}^{\infty} \int_{-\infty}^{\infty} E_R(x,y) h(x,y) \exp\left[-i\frac{\pi}{\lambda d}\left(x^2 + y^2\right)\right] \exp\left[i\frac{2\pi}{\lambda d}\left(x\xi + y\eta\right)\right] dx dy \qquad (1)$$

By applying the following substitutions to the Fresnel transform in Eq. (1)

$$\nu = \frac{\xi}{\lambda d}, \mu = \frac{\eta}{\lambda d} \qquad (2)$$

Here with (1) becomes:

$$\Gamma(\nu,\mu) = \frac{i}{\lambda d}\exp\left(\frac{-i2\pi d}{\lambda}\right)\exp\left[-i\pi\lambda d\left(\nu^2+\mu^2\right)\right]$$
$$\times \int_{-\infty}^{\infty}\int_{-\infty}^{\infty} E_R(x,y)h(x,y)\exp\left[-i\frac{\pi}{\lambda d}\left(x^2+y^2\right)\right]\exp\left[i2\pi(x\nu+y\mu)\right]dxdy \quad (3)$$

A simplified formula for reconstructing the complex amplitude of the object wavefront is obtained and expressed through the reconstructed wavefront.

$$\Gamma(\nu,\mu) = \frac{i}{\lambda d}\exp\left(\frac{-i2\pi d}{\lambda}\right)\exp\left[-i\pi\lambda d\left(\nu^2+\mu^2\right)\right]$$
$$\times \mathfrak{I}^{-1}\left\{E_R(x,y)h(x,y)\exp\left[-i\frac{\pi}{\lambda d}\left(x^2+y^2\right)\right]\right\} \quad (4)$$

Where $\mathfrak{I}^{-1}$ is the inverse Fourier transform.

In the presented configuration of the holographic setup, both the object and the point source which is emitting a spherical wave, are positioned within the same plane. Consequently, the spherical phase factor related to the Fresnel diffraction of the transmitted wave through the hologram is eliminated. This elimination is achieved through the utilization of a spherical reference wave, denoted as $E_R(x,y)$ possessing the identical average curvature as employed during the recording process, which can be expressed as follows:

$$E_R(x,y) = (const.)\exp\left(i\frac{\pi}{\lambda d}\left(x^2+y^2\right)\right) \quad (5)$$

By substituting $E_R(x,y)$ into Eq. (4), we derive the formula for reconstructing the complex amplitude of the object wavefront, expressed as:

$$\Gamma(\nu,\mu) = \frac{i}{\lambda d}\exp\left(\frac{-i2\pi d}{\lambda}\right)\exp\left[-i\pi\lambda d\left(\nu^2+\mu^2\right)\right]\mathfrak{I}^{-1}\{h(x,y)\} \quad (6)$$

In Eq.(6), inverse fast Fourier transform (IFFT) of only single term, that is, recorded digital hologram is evaluated apart from some multiplicative constant. Thus, this method is simpler and faster as compared to other reconstruction methods such as Fresnel reconstruction and convolution method, in which combination of several Fourier transform and complex multiplications need to be evaluated. The simpler and faster reconstruction algorithm enhances the possibility of real-time applications [16].

For HI measurements, the phase can be calculated from the complex amplitude as:

$$\phi_1(\xi,\eta) = \tan^{-1}\left[\frac{\text{Im}[\Gamma_1(\xi,\eta)]}{\text{Re}[\Gamma_1(\xi,\eta)]}\right] \quad (7a)$$

$$\phi_2(\xi,\eta) = \tan^{-1}\left[\frac{\text{Im}[\Gamma_2(\xi,\eta)]}{\text{Re}[\Gamma_2(\xi,\eta)]}\right] \quad (7b)$$

where "Im" and "Re" represents the imaginary and real part of the reconstructed complex amplitude of the wave.

Considering that the calculation of (2) yields values in the range [-π, π], the wrapped interference phase map (interferogram) can be achieved by performing the following operation [14]

$$\Delta\phi(\xi,\eta) = \phi_2(\xi,\eta) - \phi_1(\xi,\eta) \tag{8}$$

The subsequent unwrapping processes are generally needed to describe the actual object deformation or displacement with respect to its original state [15]. The flowchart depicted in Fig. 1 provides a visual representation of the sequential steps in the data processing and analysis procedure.

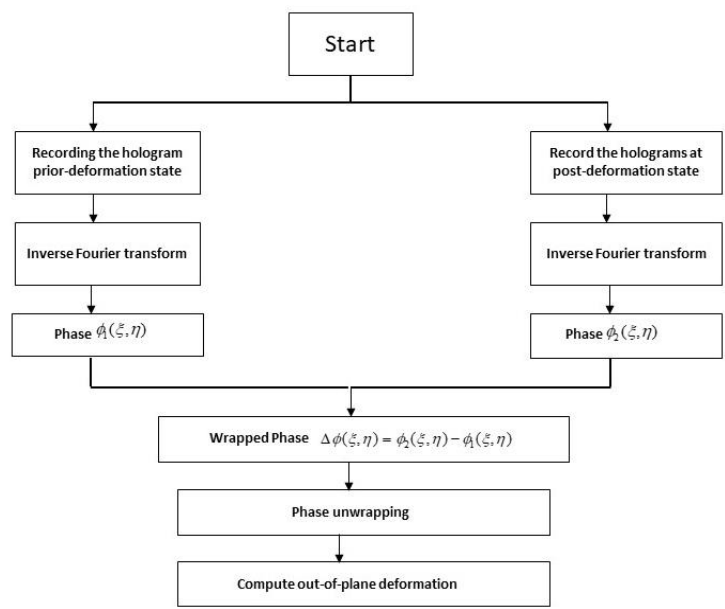

Figure 1: Flowchart of the reconstruction algorithm from the recorded holograms

**Experimental Results and Discussion**

For temporal deformation analysis of a test object, we used a custom-built off-axis digital holographic interferometry setup as shown in Fig. 2. In this experimental setup, a coherent light source (LS) of wavelength 632.8 nm is used. A spatial filter (SP) is used in two arms to clean and expand the beam size to illuminate the full area of the object. The light from beam expander is divided into two beams using a variable beamsplitter (VBS). The light wave directly incident on the camera is called the reference wavefield. On the other hand, another light wave is incident on the object (the butterfly wing, OB), and the scattered wave from the morpho butterfly wing reaches the camera. This wave is referred to as the object wavefield (O). Mirrors (M) are used to steer the laser beams along the paths. The object is deformed using a green diode laser (532nm) as an excitation laser as depicted in the Figure 2. Laser power value for excitation laser were in the range between 0.25mw to 2.5mw with a beam radius of about 3mm which was measured using beam profilers (Thorlab BC106-VIS - CCD Camera). The interference between the scattered object wave and reference wave on the camera generates a hologram. The object is clamped between two aluminum rings with diameter 20 mm and thickness 2 mm. Corresponding to different deformation states, a series of holograms are recorded on the camera (CMOS Atlas10 20.4 MP Model) sequentially at 20fps. The complex amplitude is obtained from a recorded hologram using a numerical reconstruction procedure as described before.

Simulations were implemented in MATLAB, following equations (1) to (3). For small deflections (in the order of a few micrometers) one can model the deformation of the butterfly wing. For simulations we have used as inputs the unwrapped phase values obtained from interferograms.

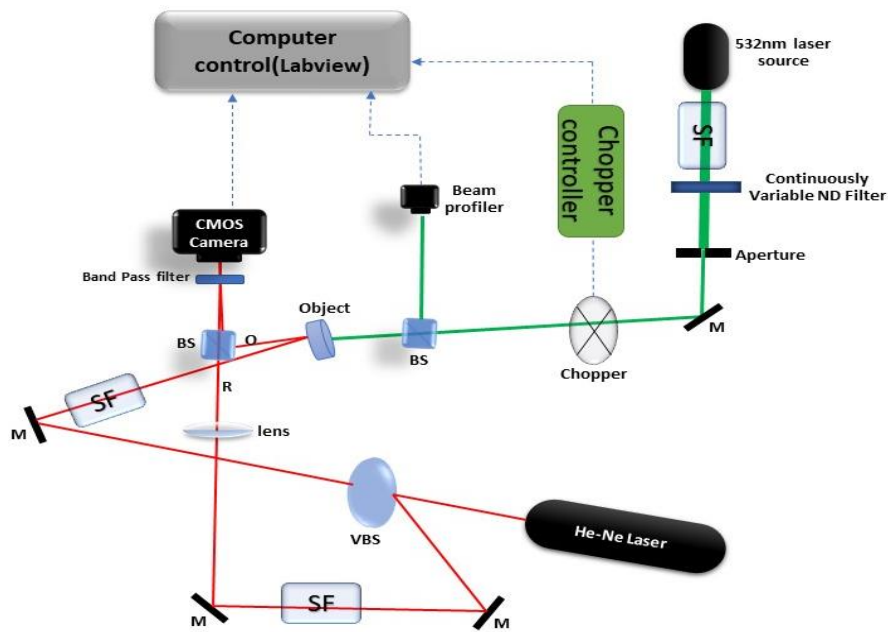

Figure 2: The schematic of setup for a digital off-axis lensless Fourier holography arrangement.(VBS) Variable beam splitter, (M) Mirror, (SF) Spatial filter, (BS) Beam Splitter.

A selected part of the wing of a Morpho butterfly was cut using laser and mounted around a circular ring, with the non-iridescent side facing at the excitation part. The wing itself is fragile and can easily be damaged during mechanical cutting. To overcome this issue, the samples were cut using a commercial $CO_2$ laser cutter without a protective air stream. The wing was cut to avoid veins which are stiffer than the membrane, and almost completely non-responsive due to a limited number of scales covering them. The diameter of the wing was around 25mm is used as the object to validate our method experimentally. Thanks to the large field of view of the imaging system, the whole wing is adequately imaged.

Numerical processing is required to recover the phase information of the object under study from each recorded experimental hologram, Fig. 3. A smaller region, marked by red in Fig. 3, was selected as the region of interest.

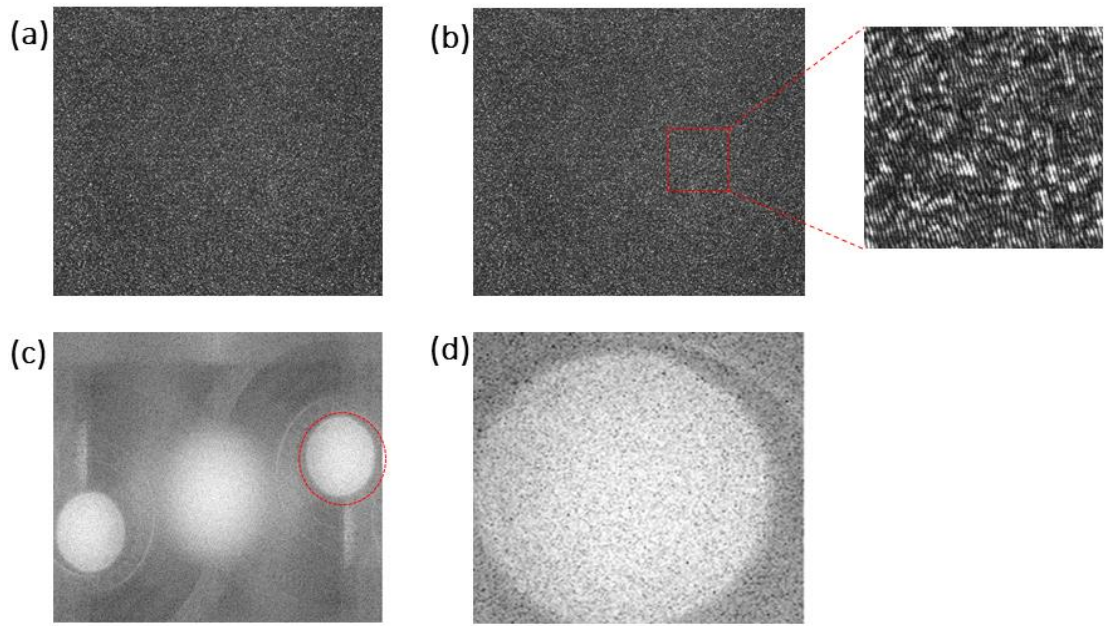

Figure 3: (a) Hologram prior to deformation. (b) Hologram after deformation. (c) intensity of complex field before deformation (d) Closer view of (c) showing the region of interest

The interference phase maps of Fig. 4 were obtained as the morpho butterfly wing was excited by the excitation laser at different intensities. To calculate phase differences, which can be later related to microdeformations of the object, a reference hologram of the object in a non-deformed state is reconstructed, and its phase recovered via Eq. (2). The pointwise $2\pi$-module subtraction between the phase information recovered from every hologram and the phase information recovered from the reference hologram yields the phase difference, Fig. 4.

**Quantitative Measurement**

In our research, we conducted a series of experiments and simulations aimed at analyzing and quantifying the subtle deflections observed in a Morpho butterfly wing caused by excitation laser. To achieve accurate measurements, we carefully arranged a holographic setup, ensuring that the object surface of the wing remained parallel to the image plane in its initial, undeformed state. This alignment allowed us to capture the wing's deflection directly in the direction of the camera while inducing excitation. Consequently, we could consider the deformation as primarily manifesting as an out-of-plane displacement. We present our findings on quantifying the out-of-plane displacement of a Morpho butterfly wing when subjected to laser excitation, [16]

$$h = \Delta\varphi \frac{\lambda}{4\pi} \qquad (9)$$

A phase variation of 2π correspond to a out-of-plane deformation of λ/2.

The camera recorded off-axis digital holograms of the wing before, during and after the excitation process, operating at a frame rate of 20(fps). By subtracting the phase of the undeformed (reference) state from that of the deformed state, we obtained the wrapped phase differences of the wing at various time intervals. The laser excitation was applied for a duration of 150 milliseconds, followed by a period of 450 milliseconds with no laser excitation.

The interference phase maps depicted in Figure.4 (a)to(d) were acquired for different laser excitation intensities falling on the wing were obtained as the morpho wing was excited by increasing intensities from 14.1 mW/cm$^2$ to 70.7 mW/cm$^2$. The phase interval is from -π to π, Subsequent unwrapping procedures allowed us to derive the relative out-of-plane deformation of the wing.

Employing equation (9), we quantified the relative deflections, yielding values in the range of 1.50 µm to 2.90 µm for the maximum deflection amplitudes of the wing. By maintaining a consistent power level for each excitation cycle, we ensured that the estimated relative errors stayed below 10%, thus confirming the precision of our quantifications. The average relaxation process for the wing samples, as observed in our investigation, lasted around 350 milliseconds. This is distinctly evident in Figure 4, where recordings were made using a camera set at 20 frames per second, with a 150-millisecond exposure on the wing followed by a subsequent 450-millisecond off period.

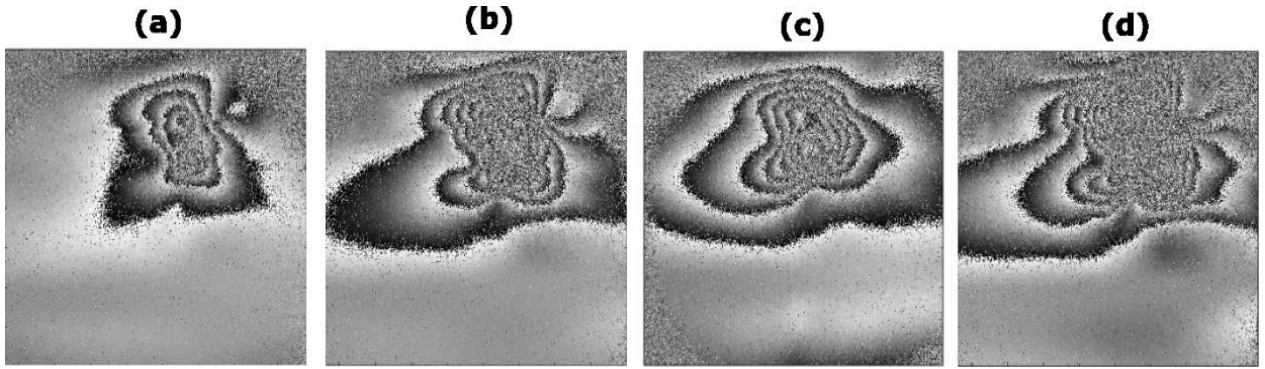

Figure 4: wrapped phase differences obtained for each laser intensity. (a) 14.1 mW/cm$^2$ (b) 25.4 mW/cm$^2$ (c) 42.4 mW/cm$^2$ (d) 70.7 mW/cm$^2$

To visualize the changes, Fig. 5 presents the wrapped phase differences. Additionally, the corresponding unwrapped phase maps are displayed. This depiction allows us to observe the progression of phase differences.

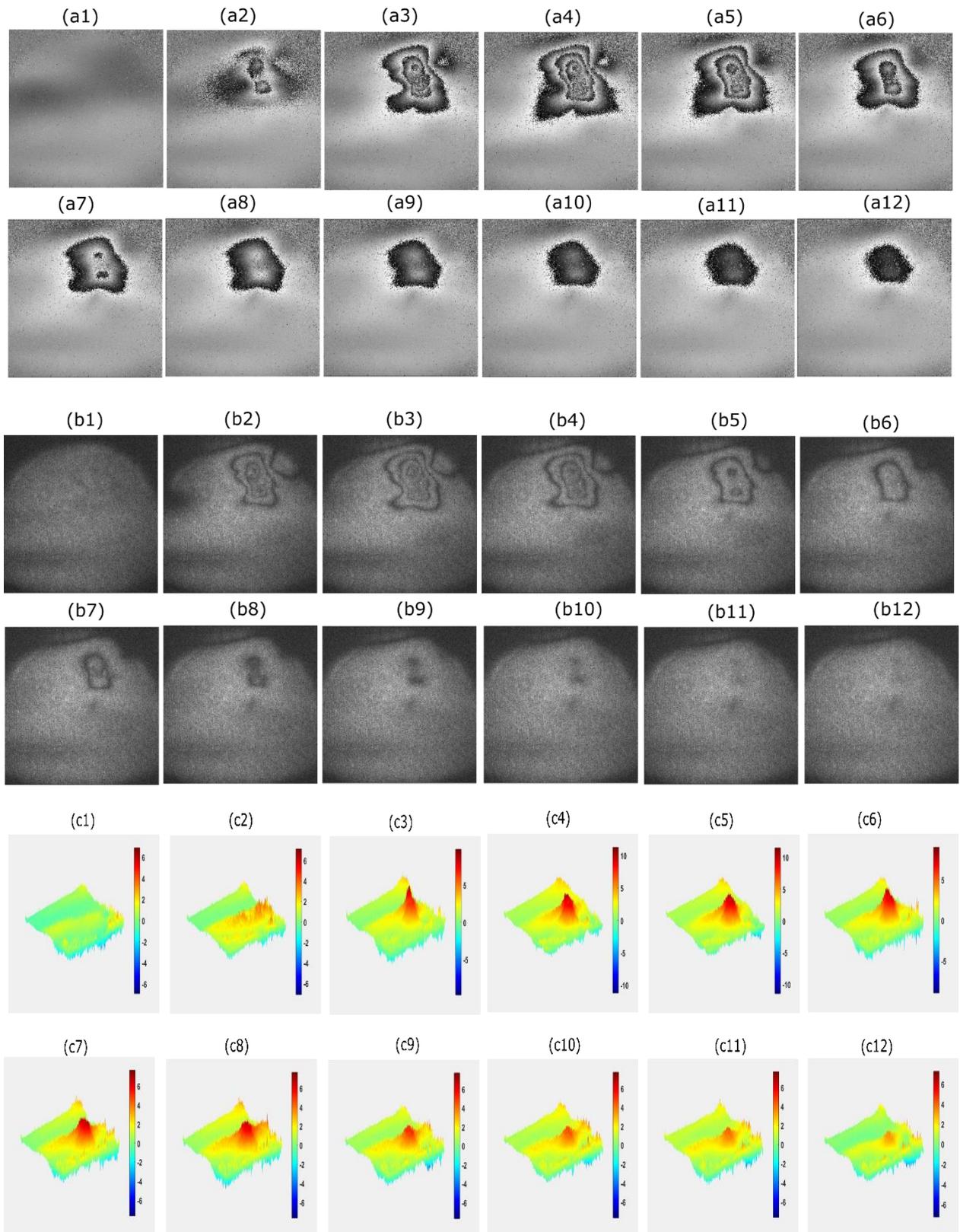

Figure 5: Dynamic displacement of the wing during excitation by 14.1 mW/cm$^2$ laser intensities is illustrated in (a1) to (a12) wrapped phase differences, (b1) to (b12) amplitude reconstruction and (c1) to (c12) the corresponding unwrapped phase.

The interference phase maps of Fig. 5(a1) to (a12) were obtained as the morpho wing was excited by laser intensity 14 mW/cm$^2$. Subsequent unwrapping procedures yield the actual deformation of the wing. By employing (9) we could quantify the deflections, obtaining values in the range between 1.5 μm to 2.9 μm for the maximum deflection amplitudes of the wing.

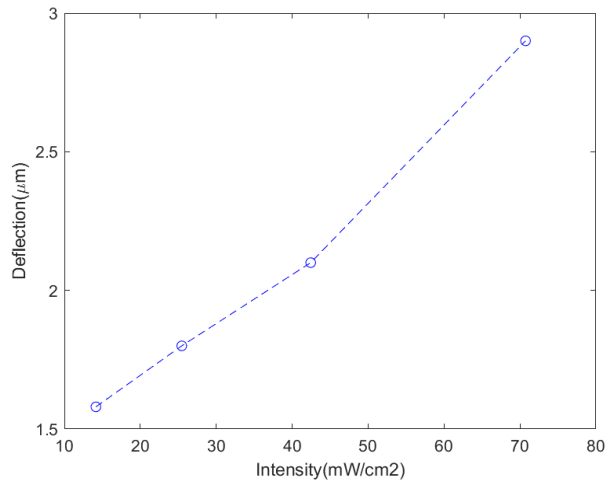

Figure 6: The physical deflection amplitudes as calculated from (4)

We would like to make a brief note that when using a low intensity of 7.0 mW/cm$^2$, we did not detect any fringes in the phase maps. This absence of fringes can be regarded as the method's operational limit. The physical deflection amplitudes were obtained by taking the profiles from the unwrapped phase maps. We observed a linear deflection pattern regarding the excitation power.

As already mentioned, for the numerical simulations we took as inputs the phase values of the maximum out-of-plane deflections calculated from unwrapping the phase maps of Fig. 5. Simulations proceeded in a rather simple way and were based on determining the value of phase for each point of the object.

The method we've introduced in this study has a broad range of applications for investigating minor deformations in sensitive objects. In this research, we deliberately selected a relatively complicated object for its unpredictable results. However, it's important to note that this method can be employed for analyzing simpler objects. As the complexity of the object decreases, so does the challenge of analyzing the collected data.

**CONCLUSION**

In conclusion, comprehending the interplay between phase information, mechanical deformation, and radiation exposure in wing scales is significant for advancing radiation detection technologies. The ability of wing scales to respond to radiation-induced changes by altering their phase information and mechanical properties underscores their potential as sensitive and effective detectors in the field of radiation monitoring. Further research in this area holds promise for developing innovative and efficient radiation detection systems inspired by nature's intricate designs.

We conducted an analysis of small deflections using digital holographic interferometry. Our approach involved employing an off-axis lensless Fourier holographic setup to observe the deflection of a morpho butterfly wing subjected to laser excitation as the heating. We recorded holograms at various excitation power level. Subsequently, we applied the Fresnel method to reconstruct the holograms and extracted phase-related information, enabling us to calculate phase differences. This process generated interferograms that illustrated the displacement of the wing between its reference (undeformed) state and excited (deformed) states.

By applying unwrapping techniques, we were able to reconstruct the actual out-of-plane deformation of the wing and determine the deflection amplitude. Assuming out-of-plane deformation, we quantified it by considering the sensitivity vector, which is solely dependent on the holographic arrangement. The outcomes of this study suggest promising avenues for further research in the development of highly precise sensing devices, particularly in the domain of non-contact metrology using digital holography. The application of advanced digital holographic interferometry opens up new possibilities for high-resolution and non-invasive measurement techniques, which are important in fields such as optical metrology, vibroacoustics, thermal analysis, and surface characterization. These findings advocate for the continued exploration and refinement of digital holography methodologies, aiming to enhance their efficacy in a broader spectrum of precision measurement and sensing applications.

**Funding.** Croatian Science Foundation (DOK-2020-01-8574), NATO SPS MYP G5618, Ministry of Science and Eduation of Republic of Croatia grant No. KK.01.1.1.01.0001.

**Data Availability**

The data that support the findings of this study are available from the corresponding author upon reasonable request.

**Disclosures**

The authors declare no conflicts of interest.